\documentclass[twocolumn,prc,aps,floatfix]{revtex4}
\usepackage[dvips]{graphicx}
\begin{document}
\newcommand{\be}{\begin{eqnarray}}
\newcommand{\ee}{\end{eqnarray}}
\title{Variation of fundamental constants: theory and observations}
\author{V.V. Flambaum}
\affiliation{
 School of Physics, The University of New South Wales, Sydney NSW
2052, Australia\\ and\\ Institute for Advanced Study,
Massey University (Albany Campus),\\Private Bag 102904, North Shore MSC
 Auckland, New Zealand}
\date{\today}
\begin{abstract}
 Review of recent works devoted to the variation
of the fine structure constant $\alpha$, strong interaction and
 fundamental masses (Higgs vacuum) is presented. The results from
 Big Bang nucleosynthesis, quasar absorption spectra,
and Oklo natural nuclear reactor data give us  the space-time
variation on the  Universe lifetime scale.
Comparison of different atomic clocks gives us the present time variation.
Assuming linear variation with time we can compare different results.
The best limit on the variation of the electron-to-proton mass ratio
$\mu=m_e/M_p$ and $X_e=m_e/\Lambda_{QCD}$ follows from the quasar
 absorption spectra \cite{FK1}:
 $\dot{\mu}/\mu=\dot{X_e}/X_e=(1 \pm 3) \times 10^{-16}$~yr$^{-1}$.
A combination of this result and the atomic clock results
\cite{clock_1,tedesco} gives the best limt on variation of $\alpha$:
$\dot{\alpha}/\alpha=(-0.8 \pm 0.8) \times 10^{-16}$~yr$^{-1}$.
The Oklo natural reactor gives the best limit on the variation
of $X_s=m_s/\Lambda_{QCD}$ where  $m_s$ is the strange quark mass
 \cite{FlambaumShuryak2002,Oklo}:
$|\dot{X_s}/X_s| < 10^{-18}$~yr$^{-1}$. Note that the Oklo
data can not give us any limit on the variation of $\alpha$
since the effect of $\alpha$ there is much smaller than the effect of $X_s$
and should be neglected. 

 Huge
 enhancement of the relative variation effects happens in transitions between
 close  atomic, molecular and nuclear energy levels.
We suggest several new cases where the levels are very  narrow.
Large enhancement of the variation effects is also possible
in cold  atomic and molecular collisions near Feshbach resonance.

How changing physical constants and violation of local
position invariance may occur? Light scalar fields very naturally
appear in modern cosmological models, affecting  parameters
of the Standard Model  (e.g. $\alpha$). Cosmological variations of
 these scalar fields should occur because of drastic changes of
 matter composition in Universe: the latest such event is rather
recent (about 5 billion years ago), from matter to dark energy domination.
Massive bodies (stars or galaxies) can also affect physical constants.
They have large scalar charge $S$ proportional to number of particles
which produces a Coulomb-like scalar field  $U=S/r$. This leads to a
variation of the fundamental constants proportional to the gravitational
potential, e.g. $\delta \alpha/ \alpha = k_\alpha \delta (GM/ r c^2)$.
We compare different manifestations of this effect.
 The strongest limits \cite{FS2007}
  $k_\alpha +0.17 k_e= (-3.5\pm 6) \times 10^{-7}$ and 
 $k_\alpha +0.13 k_q= (-1\pm 17) \times 10^{-7}$ are obtained from
 the measurements of  dependence
 of atomic frequencies on the distance from  Sun \cite{clock_1,Ashby}
 (the distance varies due
 to the ellipticity of the Earth's orbit).
\end{abstract}
\maketitle

\section{Introduction}\label{introduction}
A search for the variations of the fundamental constants is currently
 a very popular research topic.
 Theories unifying gravity and other interactions suggest
 the possibility of spatial
and temporal variation of physical ``constants'' in the Universe (see,
e.g. \cite{Marciano,Uzan}).  Moreover, there exists a
mechanism for making all coupling constants and masses of elementary
particles both space and time dependent, and influenced by local
circumstances (see e.g. review \cite{Uzan}).
  The variation of coupling constants can be
non-monotonic (for example, damped oscillations).

 These variations are usually associated
   with the effect of massless (or very light) scalar fields.
One candidate is the 
dilaton: a  scalar which appears in string theories together with a
graviton, in a massless multiplet of closed string excitations.
Other scalars naturally appear in  cosmological
models, in which our Universe is a ``brane" floating in a space
 of larger dimensions.
The scalars  are simply brane coordinates in extra dimensions.
However, the only relevant scalar field recently discovered, the cosmological
 dark energy, so far does not show  visible variations.
Available observational limits on physical constant variations at present time
  are quite strict, allowing only scalar coupling tiny in comparison with
  gravity.

 A possible explanation was suggested by Damour et al
 \cite{Damour1,Damour:1994zq} who pointed out
that  cosmological
evolution of scalars naturally leads to their self-decoupling.    
 Damour and Polyakov have further suggested
 that  variations should happen when the scalars get excited by 
some physical change in the Universe, such as the phase transitions or other
drastic change in  the equation of State of the
Universe. They considered
 few of them, but since the time of their paper a new fascinating
  transition has been discovered:
 from matter dominated (decelerating) era to dark energy 
dominated  (accelerating) era. It is relatively recent event, corresponding to
 cosmological redshift $z\approx 0.5$.    

  The time dependence of the perturbation related to it
  can be calculated, and it turned out \cite{Barrow,Olive} that the  self-decoupling
process is effective enough
to explain why after this transition the variation of constants is as small as observed in
laboratory
experiments at the present time, as well as at Oklo ($\sim 2$ billion
years ago or $z=0.14$) and isotopes ratios in meteorites ( $4.6$ billion years to now, $z=0.45-0$), while being
 at the same time consistent with possible observations of the
variations of the electromagnetic fine structure constant   at $z\sim 1$.

Another  topic we will address here  is similar variations of
constants in space,  near massive
bodies such as stars (Sun), pulsars, Galaxy. We will compare possible sensitivities
related with  different possible objects, point out limitations following
from some recent experiments with atomic clocks and suggest new measurements
( this part is based on Ref. \cite{FS2007}).

Recent observations have produced several  hints for the variation of the
fine structure constant, $\alpha=e^2/\hbar c$,  
strength constant of the strong interaction and masses
in Big Bang nucleosynthesis, quasar absorption spectra and Oklo
natural nuclear reactor data
(see e.g.\cite{Murphy,Dmitriev,Lam,Ubach}) .
However, a majority
of publications report only limits on  possible variations
 (see e.g. reviews \cite{Uzan,karshenboim}).
 A very sensitive method to study the
 variation in a laboratory
 consists of the comparison of different optical and microwave atomic clocks
(see  recent measurements in
 \cite{prestage,Marion2003,Bize2005,Peik2004,Bize2003,
Fischer2004,Peik2005,Peik2006,clock_1}).

Sensitivity to temporal variation of the fundamental constants may
be strongly enhanced in transitions between narrow close levels of
different nature. Huge enhancement of the relative variation effects 
 can be obtained
in transition between the almost degenerate levels in 
 atoms \cite{dzuba1999,Dy,nevsky,budker,budker1},
molecules \cite{DeMille,mol,ammonia,FK1,FK2} and nuclei \cite{th,th4}.

\section{Optical spectra}
\subsection{Comparison of quasar absorption spectra with laboratory spectra}
To perform measurements of $\alpha$ variation by comparison of cosmic and
 laboratory optical spectra  we developed a new approach
\cite{dzubaPRL,dzuba1999} which improves the sensitivity to a
variation of $\alpha$ by more than an order of magnitude.
  The relative value of any relativistic
corrections to atomic transition frequencies is proportional to
$\alpha^2$. These corrections can exceed the fine structure interval
between the excited levels by an order of magnitude (for example, an
$s$-wave electron does not have the spin-orbit splitting but it has the
maximal relativistic correction to energy). The relativistic corrections
vary very strongly from atom to atom and can have opposite signs in
different transitions (for example, in $s$-$p$ and $d$-$p$
transitions). Thus, any variation of $\alpha$ could be revealed by
comparing different transitions in different atoms in cosmic and laboratory
spectra.

This method provides an order of magnitude precision gain compared to
measurements of the fine structure interval.  Relativistic many-body
calculations are used to reveal the dependence of atomic frequencies on
$\alpha$ for a range of atomic species observed in quasar absorption
spectra \cite{dzuba1999,dzubaPRL,Dy,q}.  It is convenient to present results for the
transition frequencies as functions of $\alpha^2$ in the form
\begin{equation}
\label{q1}
\omega = \omega_0 + q  x,
\end{equation}
where $x = (\frac{\alpha}{\alpha_0})^2 - 1 \approx
 \frac{2 \delta \alpha}{\alpha}$ and
 $\omega_0 $ is a laboratory frequency of a particular transition.
We stress that the second  term contributes only if $\alpha$
deviates from the laboratory value $\alpha_0$. 
We performed accurate many-body calculations of the coefficients $q$
for all transtions of astrophysical interest (strong E1 transtions from the
ground state)  in Mg, Mg II, Fe II, Cr II, Ni II, Al II, Al III, Si II,
and Zn II. It is very important that this  set of transtions
contains three large classes : positive shifters (large positive coefficients
$q > 1000 $ cm$^{-1}$), negative shifters (large negative coefficients
$q <- 1000 $ cm$^{-1}$) and anchor lines with small values of $q$.
This gives us an excellent control of systematic errors
since systematic effects do not ``know'' about sign and magnitude
of $q$. Comparison of cosmic frequencies $\omega$  and
 laboratory frequencies $\omega_0$ allows us
to measure $\frac{ \delta \alpha}{\alpha}$.

 Three independent samples of data contaning 143
absorption  systems spread over red shift range $0.2 <z < 4.2$.
 The fit of the data gives \cite{Murphy}
 is $\frac{ \delta \alpha}{\alpha}=
(-0.543 \pm 0.116) \times 10^{-5}$. If one assumes the linear dependence
of $\alpha$ on time, the fit of the data gives $d\ln{\alpha}/dt=
(6.40 \pm 1.35) \times 10^{-16}$ per year
 (over time interval about 12 billion years).
 A very extensive search for possible
systematic errors has shown that known systematic effects can not explain
 the result (It is still  not completely excluded that the effect may be
 imitated by a large change of abundances of isotopes
 during last 10 billion years. 
We have checked that different isotopic abundances for any single
element can not imitate the observed effect. It may be an 
improbable ``conspiracy'' of several elements).

 Recently our method and calculations
 \cite{dzuba1999,dzubaPRL,Dy,q}
were used by two other groups \cite{chand,Levshakov}. However, they have
not detected any variation of $\alpha$. Most probably, the difference
is explained by some undiscovered systematic effects. However,
another explanation is not excluded.
   These results of \cite{Murphy} are based on the data from the
 Keck telescope which is located in the Northen hemisphere  (Hawaii).
 The results of \cite{chand,Levshakov}
are based on the data from the different telescope (VLT) located
in the Southern hemisphere (Chile). Therefore, the difference in the results
may be explained by the spatial variation of $\alpha$. 

  Recently the results of \cite{chand} were questioned in Ref.
 \cite{Murphy2006}. Re-analysis of Ref. \cite{chand}
data revealed flawed parameter estimation methods.
The authors of \cite{Murphy2006} claim that the same spectral data fitted 
more accurately give  $\frac{ \delta \alpha}{\alpha}=
(-0.44 \pm 0.16) \times 10^{-5}$ (instead of  $\frac{ \delta \alpha}{\alpha}=
(-0.06 \pm 0.06) \times 10^{-5}$ in Ref.\cite{chand}). However, even this
revised result may require further revision. 

    Using opportunity I would like to ask for new, more accurate laboratory
 measurements of  UV transition frequencies which have been observed
in the quasar absorption spectra. The ``shopping list'' is presented
in \cite{shopping}. We also need the laboratory measurements of 
isotopic shifts  - see \cite{shopping}. We have performed very complicated
calculations of these isotopic shifts \cite{isotope}. However, the
 accuracy of  these calculations in atoms and ions with open d-shell
(like Fe II, Ni II, Cr II, Mn II, Ti II) may be very low. The measurements
 for at list few lines are needed to test these calculations.
These measurements would be very important for a study of  evolution of
 isotope abundances in the Universe, to exclude the systematic effects
 in the search for 
$\alpha$ variation and to test models of nuclear reactions in stars
and supernovi.
\subsection{Optical clocks}
Optical clocks also include transitions which have positive, negative
or small constributions of the relativistic corrections to frequencies.
We used the same methods of the relativistic many-body calculations
to  calculate the dependence on $\alpha$ \cite{dzuba1999,clock,Dy}.
 The coefficients
$q$ for optical clock transitions  may be substantially larger than
in cosmic transitions since the clock transitions are often in  heavy atoms
(Hg II, Yb II, Yb III, etc.) while cosmic spectra contain mostly light
atoms lines ($Z <33$). The relativistic effects are proporitional
to $Z^2 \alpha^2$.    

\section{Enhanced effects of $\alpha$ variation in atoms}
An  enhancement of the relative effect of $\alpha$ variation can be obtained
in transition between the almost degenerate levels in Dy atom
 \cite{dzuba1999,Dy}.
These levels move in opposite directions if  $\alpha$ varies. The relative
variation may be presented as $\delta \omega/\omega=K \delta \alpha /\alpha$
 where the coefficient $K$ exceeds $10^8$. Specific
 values of $K=2 q/\omega$ are different for different 
 hyperfine components and isotopes which have different $\omega$;
 $q=30,000$ cm$^{-1}$,  $\omega \sim 10^{-4}$ cm$^{-1}$. 
 An experiment is currently underway to place limits on
$\alpha$ variation using this transition \cite{budker,budker1}.
The current limit is
 $\dot{\alpha}/\alpha=(-2.7 \pm 2.6) \times 10^{-15}$~yr$^{-1}$.
Unfortunately, one of the levels has  quite a large linewidth
and this limits the accuracy.

Several enhanced effects of $\alpha$ variation in atoms have been calculated
in \cite{nevsky}.

\section{Enhanced effects of $\alpha$ variation in molecules}
 The relative effect of $\alpha$ variation 
in microwave transitions between very close and narrow
rotational-hyperfine levels
may be enhanced 2-3 orders of magnitude in diatomic molecules
with unpaired electrons
like LaS, LaO, LuS, LuO, YbF and similar molecular ions \cite{mol}.
 The enhancement is a result
of cancellation between the hyperfine and rotational intervals; 
$\delta \omega/\omega=K \delta \alpha /\alpha$
 where the coefficients $K$ are between 10 and 1000.

 This enhancement may also exist in a large number
of molecules due to cancelation between the ground state fine
structure $\omega_f$ and vibrational interval $\omega_v$
($\omega=\omega_f-n \omega_v\approx 0$ , $\delta \omega/\omega=K (2
\delta \alpha/\alpha - 0.5 \delta \mu/\mu)$, $K \gg 1$,
$\mu=m_e/M_p$ - see \cite{FK2}). The intervals between the levels
 are conveniently
located in microwave frequency range and the level widths are very
small. Required accuracy of the shift measurements is about 0.01-1
Hz. As examples, we consider molecules Cl$_2^+$, CuS, IrC, SiBr and
HfF$^+$. An enhancement due to the cancellation between the electron
and vibrational intervals in Cs$_2$ molecule was suggested earlier
by D. DeMille \cite{DeMille}.

\section{Variation  of the strong interaction}

 The hypothetical unification of all interactions implies that a variation
in $\alpha$ should be accompanied by a variation of the strong interaction
strength and the fundamental masses. For example, the grand unification models
discussed in Ref. \cite{Marciano} predicts the quantum chromodynamics
(QCD) scale $\Lambda_{QCD}$ (defined as the position of the Landau pole in
the logarithm for the running strong coupling constant,
$\alpha_s(r) \sim 1/\ln{(\Lambda_{QCD} r/\hbar c)}$) is modified as
 ${\delta\Lambda_{QCD}}/{\Lambda_{QCD}} \approx 34 \;{\delta\alpha}/{\alpha}$.
The variations of quark mass $m_q$ and electron masses $m_e$
 ( related
to  variation of the  Higgs vaccuum field which generates fundamental
 masses)  in this model are given by 
${\delta m}/{m} \sim 70 \; {\delta\alpha}/{\alpha} $, giving an estimate of
 the variation for the dimensionless ratio
\begin{equation}\label{eq:alpha}
\frac{\delta (m/\Lambda_{QCD})}{(m/\Lambda_{QCD})} \sim
 35\frac{\delta\alpha}{\alpha}
\end{equation}
The coefficient here is model dependent but large values are generic
 for grand unification models in which modifications come from high energy
 scales; they appear because the running strong-coupling constant and
 Higgs constants  (related to mass) run faster than $\alpha$.

  Indeed, the  strong (i=3), 
and electroweak  (i=1,2) inverse coupling constants have the following
 dependence on the scale $\nu$ and normalization point $\nu_0$:
\begin{equation} \label{eqn_inv_alpha}
\alpha_i^{-1}(\nu)=\alpha_i^{-1}(\nu_0)+b_i ln(\nu/\nu_0)
\end{equation}
In the Standard Model $2\pi b_i=41/10,-19/6,-7$ and the couplings are
related as $\alpha^{-1}=(5/3)\alpha_1^{-1}+\alpha_2^{-1}$.
There are two popular scenarios of Grand Unification: with the standard model  as well as
for its minimal supersymmetric extension (MSSM). In the latter case 
 3 curves  for $\alpha_i$ (i=1,2,3) cross
at one point, believed to be a ``root" of the three branches (electromagnetic, weak and strong). One may select the unification point for $\nu_0$, and
for example, $\nu=m_Z$ is the $Z$-boson mass ( String theories lead to more
 complicated ``trees",
which however also have a singly ``root", at a string scale $\Lambda_s$
 and bare string coupling $g_s$.)

Basically there are two possibilities.
If one assumes that only $\alpha_{GUT}\equiv \alpha_i(\nu_0)$ varies, the eqn (\ref{eqn_inv_alpha})
gives us  the same shifts for all
inverse couplings
\begin{equation}
 \delta \alpha_1^{-1}=\delta \alpha_2^{-1}= \delta \alpha_3^{-1}=
 \delta \alpha_{GUT}^{-1}
\end{equation}
If so, the variation of the
 strong interaction constant $\alpha_3(m_z)$ is much larger than the
 variation
of the em constant $\alpha$, $\delta \alpha_3/\alpha_3=(\alpha_3/\alpha_1)\delta  \alpha_1/\alpha_1$.

Another option is the variation of the GUT scale ($\nu/\nu_0$ in eqn (\ref{eqn_inv_alpha})).  If so,
quite different relations between variations of the three coupling  follows
\begin{equation} \delta \alpha_1^{-1}/b_1=\delta \alpha_2^{-1}/b_2= \delta \alpha_3^{-1}/b_3 \end{equation}
Note that now variations have different sign since the one loop coefficients
$b_i$ have different sign for 1 and 2,3. Another unclear issue is the
modification of lepton/quark masses,
 which are proportional to Higgs vacuum expectation value 
and thus depend on the mechanism of electroweak symmetry breaking.

 If these models
 are correct, the variation in electron or quark masses and
 the strong interaction scale
 may be easier to detect than a variation in $\alpha$. One can only measure
 the variation of dimensionless quantities.
 The variation of $m_q/\Lambda_{QCD}$
 can be extracted from consideration of Big  Band nucleosynthesis,
 quasar absorption spectra and the Oklo natural nuclear reactor, which was
 active about 1.8 billion years ago \cite{FlambaumShuryak2002}.
 There are some
 hints for the variation in Big Bang Nucleosynthesis
 ($\sim 10^{-3}$ - see Ref.\cite{Dmitriev}) and
Oklo ($\sim 10^{-9}$ - see Ref.\cite{Lam}) data.
However, these results are not confirmed by new studies \cite{BBN,Oklo}.

   The results from Oklo natural nuclear reactor are based on the measurement
of the position of very low energy resonance ($E_r=0.1$ eV) in neutron 
capture by $^{149}$Sm nucleus. The estimate of the shift of this resonance
 induced by the
 variation of $\alpha$ have been done long time ago in works \cite{Dyson}.
Recently we performed a rough estimate of the effect of the variation of
  $m_q/\Lambda_{QCD}$ \cite{FlambaumShuryak2002}. The final result is
 \begin{equation}\label{deltaE}
\delta E_r \approx 10^6 eV (
\frac{\delta \alpha}{\alpha} -10 \frac{\delta X_q}{X_q }
+ 100 {\delta X_s \over X_s})
\end{equation} 
where $X_q=m_q/\Lambda_{QCD}$, $X_s=m_s/\Lambda_{QCD}$,
$m_q=(m_u+m_d)/2$ and $m_s$ is the strange quark mass. Refs. \cite{Oklo}
found that $|\delta E_r| < 0.1$ eV. This gives us a limit 
 \begin{equation}\label{Oklolimit}
|0.01\frac{\delta \alpha}{\alpha} -0.1 \frac{\delta X_q}{X_q }
 +{\delta X_s \over X_s}|<10^{-9}
\end{equation} 
The contribution of the $\alpha$ variation in this equation is very small
and should be neglected since the accuracy of the calculation of the main term
 is low.
Thus, the Oklo data can not give any limit on the variation of $\alpha$.
Assuming linear time dependence during last 2 billion years 
we obtain an estimate
$|\dot{X_s}/X_s| < 10^{-18}$~yr$^{-1}$.

 The proton mass is proportional to $\Lambda_{QCD}$
 ($M_p \sim 3 \Lambda_{QCD}$),
therefore, the measurements of the variation of the electron-to-proton
mass ratio $\mu=m_e/M_p$ is equivalent to the measurements of the variation of 
$X_e=m_e/\Lambda_{QCD}$. Two new results have been obtained recently
using quasar absorption spectra. In our  paper \cite{tzana}
 the varition of the
ratio of the hydrogen hyperfine frequency to optical frequencies in ions have
been measured. The result is consistent with no variation of
 $X_e=m_e/\Lambda_{QCD}$. However, in the  recent
paper \cite{Ubach} the variation
 was detected
at the level of 4 standard deviations: $\frac{ \delta X_e}{X_e}=
\frac{ \delta \mu}{\mu}=
(-2.4 \pm 0.6) \times 10^{-5}$. This result is based on the  hydrogen molecule
 spectra. Note, however, that the difference between the zero result
 of \cite{tzana} and non-zero result of \cite{Ubach} may be explained by
 a space-time variation of $X_e$. The variation of $X_e$ in \cite{Ubach}
is substantially larger than the variation of $\alpha$ measured in
 \cite{Murphy, chand}. This may be considered as an argument in favour
of Grand Unification theories of the variation \cite{Marciano}.

Recently we obtained the limit on the space-time variation of the ratio of the
proton mass to the electron mass based on comparison
of quasar absorption spectra of  NH$_3$ with CO, HCO$^+$ and HCN
rotational spectra \cite{FK1}. For the inversion transition in  NH$_3$
($\lambda\approx 1.25$~cm$^{-1}$) the relative frequency shift is
significantly enhanced: $\delta \omega/\omega=4.46\, \delta
\mu/\mu$. This enhancement allows one to increase sensitivity to the
variation of $\mu$ using NH$_3$ spectra for high redshift objects.
We use published data on microwave spectra of the object B0218+357
to place the limit  $\delta \mu/\mu =(-0.6\pm 1.9)\times 10^{-6}$ at
redshift $z=0.6847$; this limit is  several times better than the
limits obtained by different methods and may be significantly
improved.
 Assuming linear time dependence we obtain \cite{FK1}
$\dot{\mu}/\mu=\dot{X_e}/X_e=(1 \pm 3) \times 10^{-16}$~yr$^{-1}$.

\section{Microwave clocks}
 Karshenboim
 \cite{Karshenboim2000} has pointed out that measurements of ratios
 of hyperfine structure intervals in different atoms are sensitive to
 variations in nuclear magnetic moments. However, the magnetic moments
are not the fundamental parameters and can not be directly compared with
 any theory of the variations. Atomic and nuclear calculations are needed 
for the interpretation of the measurements. We have performed both
atomic calculations of $\alpha$ dependence \cite{dzuba1999,clock,Dy} and
 nuclear calculations of $X_q=m_q/\Lambda_{QCD}$ dependence \cite{tedesco}
 for all microwave transitions of current experimental interest including
hyperfine transitions in $^{133}$Cs, $^{87}$Rb, $^{171}$Yb$^+$,
$^{199}$Hg$^+$, $^{111}$Cd, $^{129}$Xe, $^{139}$La, $^{1}$H,  $^{2}$H and
 $^{3}$He. The  results for the dependence of the transition frequencies
 on variation
of $\alpha$, $X_e=m_e/\Lambda_{QCD}$ and  $X_q=m_q/\Lambda_{QCD}$
 are presented in Ref.\cite{tedesco} (see the final results in the Table IV
 of Ref.\cite{tedesco}). Also, one can find there experimental
 limits on these variations  which follow from the recent measurements. 
The accuracy is approaching $10^{-15}$ per year. This may be compared
to the sensitivity $\sim 10^{-5}-10^{-6}$ per $10^{10}$ years obtained using 
the quasar absorption spectra.

    According to Ref. \cite{tedesco} the frequency ratio $Y$ of the 282-nm
 $^{199}$Hg$^+$ optical clock transition to the ground state hyperfine
 transition
in  $^{133}$Cs has the following dependence on the fundamental constants:
\begin{equation}\label{Hg}
\dot{Y}/Y=-6\dot{\alpha}/\alpha -\dot{\mu}/\mu -0.01 \dot{X_q}/X_q
\end{equation}
In the work \cite{clock_1} this ratio has been measured:
$\dot{Y}/Y=(0.37 \pm 0.39) \times 10^{-15}$~yr$^{-1}$.
Assuming linear time dependence we obtained the quasar result \cite{FK1}
$\dot{\mu}/\mu=\dot{X_e}/X_e=(1 \pm 3) \times 10^{-16}$~yr$^{-1}$.
A combination of this result and the atomic clock result \cite{clock_1} for $Y$
 gives the best limt on the variation of $\alpha$:
$\dot{\alpha}/\alpha=(-0.8 \pm 0.8) \times 10^{-16}$~yr$^{-1}$.
Here we neglected the small ($\sim 1\%$) contribution of $X_q$.

\section{Enhanced effect of  variation of $\alpha$
and strong interaction in UV transition of $^{229}$Th nucleus (nuclear clock)}

A very  narrow level  $(3.5\pm 1)$ eV above the ground state exists
 in $^{229}$Th nucleus \cite{th1}
 (in \cite{th6} the energy is $(5.5\pm 1)$ eV,
 in \cite{th7} the energy is $(7.6\pm 0.5)$ eV). The position
of this level was determined from the energy differences of many high-energy
$\gamma$-transitions (between 25 and 320 KeV) to the ground and excited
 states. The subtraction  produces
the large  uncertainty in the position of the 3.5 eV excited state.
 The width of this level is estimated to be
about $10^{-4}$ Hz \cite{th2}. This would explain why it is so hard to find
 the direct radiation in this very weak  transition. The direct
 measurements have
 only given experimental limits on the width and energy  of this transition
 (see e.g. \cite{th3}). A detailed discussion of the measurements
 (including several unconfirmed
 claims of the  detection of the direct radiation ) is presented
 in Ref.\cite{th2}. However, the search for the direct radiation continues
\cite{private}. 

  The  $^{229}$Th transition is very narrow and can be investigated
 with laser spectroscopy.
 This makes $^{229}$Th a possible reference for an
 optical clock of very high accuracy, and opens a new possibility
for a laboratory search for the varitation of the fundamental constants
\cite{th4}.

As it is shown in Ref. \cite{th} there is an additional very important
 advantage.
The relative effects of variation of
 $\alpha$ and $m_{q}/\Lambda_{QCD}$ are enhanced by 5 orders of magnitude.
 A rough estimate for the relative variation of the $^{229}$Th
 transition frequency  is
 \begin{equation}\label{deltaf}
\frac{\delta \omega}{\omega} \approx 10^5 (
2 \frac{\delta \alpha}{\alpha} + 0.5  \frac{\delta X_q}{X_q }
 - 5 {\delta X_s \over X_s})\frac{7\,eV }{\omega}
\end{equation} 
where $X_q=m_q/\Lambda_{QCD}$, $X_s=m_s/\Lambda_{QCD}$,
$m_q=(m_u+m_d)/2$ and $m_s$ is the strange quark mass.
Therefore, the Th  experiment would
have the potential of improving the  sensitivity to temporal
variation of the fundamental
constants by many orders of magnitude.

Note that there are other narrow low-energy levels in nuclei,
 e.g. 76 eV level in $^{235}U$ with the 26.6 minutes lifetime
 (see e.g.\cite{th4}). One may expect a similar  enhancement there.
Unfortunetely, this level can not be reached with usual lasers. In principle,
 it may be investigated using a free-electron laser or synchrotron radiation.
However, the accuracy
of the frequency measurements is much lower in this case.

\section{Enhancement of variation of
fundamental constants in ultracold atom and molecule systems near
Feshbach resonances}
Scattering length $A$, which can be measured in Bose-Einstein condensate
and Feshbach molecule experiments, is extremely sensitive to the
variation of the
electron-to-proton mass ratio $\mu=m_e/m_p$ or $X_e=m_e/\Lambda_{QCD}$
 \cite{chin}.
\begin{equation}\label{d_a_final}
\frac{\delta A}{A}=K\frac{\delta\mu}{\mu}=K\frac{\delta X_e}{X_e},
\end{equation}
 where $K$ is the  enhancement factor.
For example, for Cs-Cs collisions we obtained
 $K\sim 400$. With the Feshbach resonance, however, one is
given the flexibility to adjust position of the resonance using
external  fields.  Near a narrow magnetic or an optical Feshbach resonance
 the enhancement factor $K$ may be increased by many orders of magnitude. 

\section{Changing physics near massive bodies}
In this section I follow Ref. \cite{FS2007}.

The reason gravity is so important at large scales is that
its effect is additive. The same should be true for
 massless (or very light) scalars: its effect near large body is proportional to the
 number of particles in it.
  
For not-too-relativistic objects, like the usual stars or planets,
 both their total  
mass $M$ and the total scalar charge $Q$ are simply proportional to
the number of nucleons in them, and thus the scalar field is simply
proportional to the gravitational potential
\begin{equation}\label{kappa}
 \phi-\phi_0=\kappa (GM/rc^2) \,. 
\end{equation}
Therefore, we expect that the fundamental constants would also depend on
the
position via the gravitational potential at the the measurement point.

 Naively, one may think that the larger is the dimensionless
gravity 
potential $(GM/rc^2)$ of the object
considered, the better. However, different objects allow for quite
different accuracy. 

Let us mention few possibilities, using as a comparison parameter
the product of gravity potential divided by the tentative relative accuracy
\begin{equation}
 P= (GM/rc^2)/(accuracy)
\end{equation}

(i) Gravity potential on Earth is changing due to ellipticity of its
orbit:
the corresponding variation of the Sun graviational potential is
 $\delta (GM/rc^2)=3.3 \cdot 10^{-10}$. The
accuracy of atomic clocks in laboratory conditions approaches
$10^{-16}$,
and so $P\sim 3 \cdot 10^6$. However, comparing  clocks on Earth and distant
satellite one may get variation of the Earth
 graviational potential $\delta (GM/rc^2)\sim 10^{-9}$ and $P\sim 10^7$.
The space mission was recently discussed, e.g. in the proposal 
\cite{schiller} and references therein. Note that the matter composition
of Earth and Sun is very different, therefore, the proportionality
coefficients $\kappa$ in Eq (\ref{kappa}) may also be different.
Indeed, the  first example (Sun)
is mainly sensitive to the scalar  potentials of electrons and protons
while the second example (Earth) is in addition sensitive to the scalar
  potentials of neutrons and virtual mesons mediating the nuclear forces
(the nuclear binding energy).  

(ii) Sun (or other ordinary stars) has $GM/rc^2\sim
2 \cdot 10^{-7}$. Assuming accuracy $10^{-8}$ in the measurements of
atomic spectra near the surface we get $P\sim 10$.
However, a mission with modern atomic clocks sent to the Sun would have
$P\sim 10^8$ or so, see details in the proposal \cite{Maleki}.

(iii) The stars at different positions inside our (or other) Galaxy
 have gravitational potential difference of the order of $10^{-7}$,
and (like for the Sun edge) one would expect $P\sim 10$.
Clouds which give the observable absorption lines in quasar spectra
have also different gravitational potentials (relative to Earth), 
of comparable magnitude.  

(iv) White/brown dwarfs have $GM/rc^2\sim 3 \cdot 10^{-4}$, and in some
cases
rather low  temperature. We thus get $P\sim 3 \cdot 10^4$.

(v) Neutron stars have very large gravitational potential
$GM/rc^2\sim 0.1$, but high temperature and magnetic fields make
accuracy of atomic spectroscopy rather problematic, we give tentative
accuracy 1 percent. $P\sim 10$.

(vi) Black holes, in spite of its large gravitational potential, have
no scalar field outside the Shwartzschield  radius, and thus are not 
    useful for our purpose.

Accuracy of the atomic clocks is so high because they use extremely
narrow lines. At this stage, therefore, star spectroscopy seem not to
be
competitive: the situation may change if narrow lines be identified.

Now let us see what is the best  limit available today. As an example we consider recent
work \cite{clock_1} who obtained the following value for the half-year
variation
of the frequency ratio of two atomic clocks: (i) optical transitions in 
mercury ions $^{199}Hg^+$ and (ii) hyperfine splitting
in $^{133}Cs$ (the frequency standard). The limit obtained is
\begin{equation}
 \delta ln({\omega_{Hg}\over \omega_{Cs}})=(0.7\pm 1.2) \cdot 10^{-15}
\end{equation}  
For  Cs/Hg  frequency ratio of these clocks  the dependence on the fundamental
constants
was evaluated in \cite{tedesco} with the result
\begin{equation}
 \delta ln({\omega_{Hg}\over \omega_{Cs}})=-6 {\delta \alpha \over
\alpha} -0.01{\delta  (m_q/\Lambda_{QCD}) \over (m_q/\Lambda_{QCD})} -{\delta  (m_e/M_p) \over (m_e/M_p)}
\end{equation}
Another work \cite{BW} compare $H$ and $^{133}Cs$ hyperfine transitions.
The amplitude of the half-year variation  found were
\begin{equation}
 |\delta ln(\omega_{H}/\omega_{Cs})| <7 \cdot 10^{-15}
\end{equation}
The sensitivity \cite{tedesco}
\begin{equation}
\delta ln({\omega_{H}\over \omega_{Cs}})=-0.83 {\delta \alpha \over
\alpha} -0.11{\delta  (m_q/\Lambda_{QCD}) \over (m_q/\Lambda_{QCD})}
\end{equation}
There is no sensitivity to $m_e/M_p$ because
they are both hyperfine transitions.

As motivated above, we assume that scalar and gravitational
potentials are proportional to each other,  and thus introduce parameters
$k_i$ as follows
\begin{equation}
 {\delta \alpha \over \alpha } = k_\alpha \delta ({GM\over r c^2})
\end{equation}
\begin{equation}
  {\delta (m_q/\Lambda_{QCD}) \over (m_q/\Lambda_{QCD})} = k_q \delta ({GM\over r c^2})
\end{equation}
\begin{equation}
  {\delta (m_e/\Lambda_{QCD}) \over (m_e/\Lambda_{QCD}) } = {\delta (m_e/M_p) \over (m_e/M_p) } =k_e \delta ({GM\over r c^2})
\end{equation}
where in the r.h.s. stands half-year variation of Sun's gravitational potential
on Earth. 

In such terms, the results of  Cs/Hg  frequency ratio measurement
\cite{clock_1}  can be rewritten as
\begin{equation}
 k_\alpha +0.17 k_e= (-3.5\pm 6) \cdot 10^{-7}
\end{equation}
 The results of  Cs/H  frequency ratio measurement \cite{BW} can be presented as 
\begin{equation}
| k_\alpha +  0.13 k_q | <2.5 \cdot 10^{-5}
\end{equation}
Finally, the result of recent measurement  \cite{Ashby} of
  Cs/H  frequency ratio can be presented as 
\begin{equation}
 k_\alpha +0.13 k_q= (-1\pm 17) \cdot 10^{-7}
\end{equation}
The sensitivity coefficients for other  clocks have been discussed above.

\section{Acknowledgments}
The author is grateful to E. Shuryak for valuable contribution to
 the part of this work describing dependence of the fundamental constants
 on the gravitational potential, and to D. Budker and S. Schiller for useful
 comments. 
 This work is supported by the Australian
 Research Council.


\end{document}